\documentstyle[prl, multicol, aps, epsf]{revtex}

\begin{document}
\draft

\title
{\large \bf Charge Modulations in the Superconducting State of the Cuprates}
\author{Degang Zhang}
\address{Texas Center for Superconductivity and Department of Physics,
University of Houston, Houston, TX 77204, USA}

\maketitle

\begin{abstract}

Motivated by the recent scanning tunneling microscopy (STM) and neutron
scattering experiments, we investigate various charge density wave orders
coexisting with superconductivity in the cuprate superconductors.
The explicit expressions of the local density of states and its Fourier
component at the ordering wavevector for the weak charge modulations
are derived. It is shown that the
STM experiments in $Bi_2Sr_2CaCu_2O_{8+\delta}$ cannot be explained by
a site- or bond-centered charge modulation alone, but agree well
with the presence of the dimerization hopping and transverse pairing
modulations. We also calculate the spectral function for the charged
stripes, which is measured by the ARPES experiments.

\end{abstract}

\begin{multicols}{2}

The coexistence of charged stripes and superconductivity
in high-$T_c$ superconductors  has attracted a lot of both experimental and
theoretical attention recently. In a novel STM experiment, Hoffman {\it et al}
observed a four cell checkbroad local density of states (LDOS) modulation
around the cores of superconducting vortices in
$Bi_2Sr_2CaCu_2O_{8+\delta}$ by applying a magnetic field [1].
The charge  modulation occurs in the
$Cu-O$ bond direction and in the energy range $0\le E\le12 meV$.
Very recently Howald {\it et al} discovered similar charge modulation
at energy $E=25meV$ but in absence of magnetic field [2].
The neutron scattering experiments on underdoped $YBa_2Cu_3O_{6.35}$
have also shown the charge density wave order with a period of
eight lattice constants coexisting with superconductivity [3].

To explain the STM spectra in $Bi_2Sr_2CaCu_2O_{8+\delta}$,
a number of theoretical studies based on different models
have been carried out by various authors [4-10].
However, it is important to determine the correct charge
density wave orders in order to understand the origin of
the charge stripes. In Ref. [8], Podolsky {\it et al} analysed
in detail the influence of various patterns
of translational symmetry breaking on the Fourier component of
LDOS at the ordering wavevector ${\bf Q}$ by emyloping
an approximate technique. They concluded that the STM experiments [2]
are consistent with the periodic modulation in the electron
hopping. In contrast, by the numerical simulation of a d-wave superconductor
with two-dimensional site charge density wave, bond charge
density wave or pairing modulation,
the  pairing amplitude modulation comes closest to the experimental curves
of Ref. [2] [9]. Therefore, there is no consensus about
the constitution of the charged stripes.
In this paper, in order to explain well the STM experiments,
we solve strictly  the d-wave
superconductor with one-dimensional weak charge modulations.
The results show that the dimerization hopping and transverse
pairing modulations are consistent with the STM experiments.
We also discuss the ARPES experiments on the charged stripes,
which verify the existence of the hopping and pairing modulations.

We start from the  mean-field Hamiltonian of a d-wave
superconductor
$$ H_{\rm BCS}=\sum_{{\bf K} \sigma}\epsilon_{\bf K}c^\dag_{{\bf K}
     \sigma}c_{{\bf K}\sigma}+\sum_{\bf K}\Delta_{\bf K}
     (c^\dag_{{\bf K}\uparrow}c^\dag_{{-{\bf K}}\downarrow}
     +c_{-{\bf K}\downarrow}c_{{\bf K}\uparrow}),\eqno{(1)}$$
where $\epsilon_{\bf K}=t_0+t_1({\rm cos}K_x+{\rm
cos}K_y)/2+t_2{\rm cos}K_x{\rm cos}K_y+t_3({\rm cos}2K_x+{\rm
cos}2K_y)/2 +t_4({\rm cos}2K_x{\rm cos}K_y+{\rm cos}K_x{\rm
cos}2K_y)/2+t_5{\rm cos}2K_x{\rm cos}2K_y$ and $\Delta_{\bf K}=
\Delta_0({\rm cos}K_x-{\rm cos}K_y)/2$.
For nearly optimally doped $Bi_2Sr_2CaCu_2O_{8+\delta}$, $t_{0-5}=0.1305,
-0.5951, 0.1636, -0.0519, -0.1117, 0.0510 (eV)$ [11].
For underdoped $YBa_2Cu_3O_{6.35}, t_{0-2}=0.2445, -1.2000, 0.3600 $ $(eV)$
and $t_{3-5}=0.0 (eV)$ (6 percent doping).
For both superconductors, we choose $\Delta_0=
0.0400 eV$.

The charge modulations can be introduced phenomenologically into the
Hamiltonian (1) by adding the charge ordered
parameters, which have a general form
$$\begin{array}{lll}
H_{\rm C}&=&\sum_{{\bf K}\sigma}(f_{\bf K}c^\dag_{{\bf K}+{\bf Q} \sigma}
c_{{\bf K}\sigma}+f^*_{\bf K}c^\dag_{{\bf K}\sigma}c_{{\bf K}+{\bf Q}\sigma})\\
 & &+\sum_{\bf K}(g_{\bf K}c^\dag_{{\bf K}+{\bf Q}\uparrow}
 c^\dag_{-{\bf K}\downarrow}+g^*_{\bf K}c^\dag_{{\bf K}\uparrow}
 c^\dag_{-{\bf K}-{\bf Q} \downarrow}+ h. c),
\end{array}\eqno{(2)}$$
where ${\bf Q}={\rm Q}{\bf e}_x$ (${\rm Q}=\pi/2$ for $Bi_2Sr_2CaCu_2O_{8+\delta}$
and $\pi/4$ for $YBa_2Cu_3O_{6.35}$), $f_{\bf K}(\equiv f_{-{\bf K}-{\bf Q}})$
and $g_{\bf K}(\equiv g_{-{\bf K}-{\bf Q}})$ describe the
hopping modulations and pairing modulations, respectively.
We note that $f_{\bf K}=\lambda_1, \lambda_2e^{-i{\rm Q}/2},
\lambda_3{\rm cos}(K_x+{\rm Q}/2)e^{-i{\rm Q}/2}$ and $\lambda_4{\rm cos}K_y$
are the site-centered, bond-centered, longitudinal dimerization and
transverse dimerization charge modulations while
$g_{\bf K}=\lambda_5{\rm cos}(K_x+{\rm Q}/2)e^{-i{\rm Q}/2}$ and
$\lambda_6{\rm cos}K_y$ are the longitudinal and transverse
pairing modulations [8]. In this paper, we restrict our discussion to the case of
weak charge modulations, i. e. small $\lambda$'s and do not consider the
influence of the incommensurate spin orders to LDOS. In fact, the spin orders
were not observed experimentally [2].

Taking the Bogoliubov transformation
$$ \begin{array}{lll}
c_{{\bf k}+m{\bf Q}\uparrow}&=&\xi_{{\bf k} m 0}\psi_{{\bf k} m 0}
     -\xi_{{\bf k} m 1}\psi_{{\bf k} m 1}\\
c^\dag_{-{\bf k}-m{\bf Q}\downarrow}&=&\xi_{{\bf k} m 1}\psi_{{\bf k} m 0}
     +\xi_{{\bf k} m 0}\psi_{{\bf k} m 1},
\end{array}\eqno{(3)}$$
where ${\bf k}$ is restricted to the reduced Brillouin zone,
$m=0, 1, 2, \cdots, N=(2\pi/{\rm Q})-1$,
$\xi^2_{{\bf k}m\nu}=\frac{1}{2}[1+(-1)^\nu\epsilon_{{\bf k}+m{\bf Q}}/
E_{{\bf k}+m{\bf Q}}]$, $\xi_{{\bf k}m 0}\xi_{{\bf k}m 1}=
\Delta_{{\bf k}+m{\bf Q}}/2E_{{\bf k}+m{\bf Q}}$ and
$E_{{\bf k}+m{\bf Q}}=(\epsilon^2_{{\bf k}+m{\bf Q}}+
\Delta^2_{{\bf k}+m{\bf Q}})^{\frac{1}{2}}$, the total Hamiltonian
$H=H_{\rm BCS}+H_{\rm C}$ can be rewritten as
$$\begin{array}{c}
H=\sum_{{\bf k}m\nu}(-1)^\nu E_{{\bf k}+m{\bf Q}}
\psi^\dag_{{\bf k}m\nu}\psi_{{\bf k}m\nu}\\
+\sum_{{\bf k}m \nu \nu^\prime}\{[\alpha^{\nu \nu^\prime}_{m+1 m}({\bf k})
+\beta^{\nu \nu^\prime}_{m+1 m}({\bf k})]
\psi^\dag_{{\bf k}m+1\nu}\psi_{{\bf k}m\nu^\prime}
+h. c\},
\end{array}\eqno{(4)}$$
where $\alpha^{\nu \nu^\prime}_{m+1 m}({\bf k})=f_{{\bf k}+m{\bf Q}}[(-1)^{\nu+
\nu^\prime}\xi_{{\bf k}m+1\nu}\xi_{{\bf k}m\nu^\prime}-
\xi_{{\bf k}m+1\nu+1}\xi_{{\bf k}m\nu^\prime+1}]$ and
$\beta^{\nu \nu^\prime}_{m+1 m}({\bf k})=g_{{\bf k}+m{\bf Q}}[(-1)^{\nu}\xi_
{{\bf k}m+1\nu}\\
\times \xi_{{\bf k}m\nu^\prime+1}+(-1)^{\nu^\prime}
\xi_{{\bf k}m+1\nu+1}\xi_{{\bf k}m\nu^\prime}]$.

We define two-point Green's functions
$$G^{\nu\nu^\prime}_{m m^\prime}({\bf k},{\bf k}^\prime;i\omega)
=-{\cal F}<T_\tau[\psi_{{\bf k}m\nu}(\tau)
\psi^\dag_{{\bf k}^\prime m^\prime\nu^\prime}(0)]>,\eqno{(5)}$$
where ${\cal F}\phi(\tau)$ denote the Fourier transform of
${\phi}(\tau)$ in Matsubara frequencies.
Then the equations of motion for the Green's functions are
$$\begin{array}{c}
[i\omega_n-(-1)^\nu E_{{\bf k}+m{\bf Q}}]G^{\nu\nu^\prime}_
{m m^\prime}({\bf k},{\bf k^\prime};i\omega_n)\\
-\sum_{\nu^{\prime\prime}}(-1)^{\nu+\nu^{\prime\prime}}
\xi_{{\bf k}m\nu+\nu^{\prime\prime}}[f_{{\bf k}+(m-1){\bf Q}}
{\cal G}^{\nu^{\prime\prime}\nu^\prime}_{m-1 m^\prime}
({\bf k},{\bf k^\prime};i\omega_n)\\
+ f^*_{{\bf k}+m{\bf Q}}{\cal G}^{\nu^{\prime\prime}\nu^\prime}
_{m+1 m^\prime}({\bf k},{\bf k^\prime};i\omega_n)]\\
-\sum_{\nu^{\prime\prime}}(-1)^{\nu\nu^{\prime\prime}}
\xi_{{\bf k}m\nu+\nu^{\prime\prime}+1}[g_{{\bf k}+(m-1){\bf Q}}
{\cal G}^{\nu^{\prime\prime}\nu^\prime}_{m-1 m^\prime}
({\bf k},{\bf k^\prime};i\omega_n)\\
+ g^*_{{\bf k}+m{\bf Q}}{\cal G}^{\nu^{\prime\prime}\nu^\prime}
_{m+1 m^\prime}({\bf k},{\bf k^\prime};i\omega_n)]
=\delta_{{\bf k}{\bf k}^\prime}\delta_{m m^\prime}
\delta_{\nu\nu^\prime},
\end{array}\eqno{(6)}$$
where $$\begin{array}{c}
 {\cal G}^{\nu\nu^\prime}_{m m^\prime}({\bf k}, {\bf
k}^\prime;i\omega_n)=\sum_{\nu^{\prime\prime}}
(-1)^{(1-\nu)\nu^{\prime\prime}}\xi_{{\bf k}m\nu+
\nu^{\prime\prime}}\\ ~~~~~~~\times
G^{\nu^{\prime\prime}\nu^\prime}_{mm^\prime} ({\bf k},{\bf
k}^\prime;i\omega_n). \end{array}\eqno{(7)}$$ Obviously, combining
Eqs.(6) and (7), the anomalous Green's functions in Eq. (5) can be
solved by inverting a $2(N+1)\times 2(N+1)$ matrix. To this end,
we define $$\begin{array}{c} G^0_{m\nu}({\bf
k};i\omega_n)=\frac{1}{i\omega_n-(-1)^\nu E_{{\bf k}+m{\bf Q}}},\\
a_m({\bf k};i\omega_n)=\sum_\nu\xi^2_{{\bf k}m\nu} G^0_{m\nu}({\bf
k};i\omega_n),\\ b_m({\bf k};i\omega_n)=\sum_\nu(-1)^\nu\xi_{{\bf
k}m\nu} \xi_{{\bf k}m\nu+1}G^0_{m\nu}({\bf k};i\omega_n),\\
c_m({\bf k};i\omega_n)=\sum_\nu\xi^2_{{\bf k}m\nu+1}
G^0_{m\nu}({\bf k};i\omega_n),\\
g^{\nu\nu^\prime}_{mm^\prime}({\bf k},{\bf k}^\prime;i\omega_n)
=(-1)^{(1-\nu)\nu^\prime}\delta_{{\bf k}{\bf k}^\prime}
\delta_{mm^\prime}\xi_{{\bf k}m\nu+\nu^\prime}\\ ~~~~~\times G^0_
{m\nu^\prime}({\bf k};i\omega_n),\\
\end{array}\eqno{(8)}$$
and introduce the $2(N+1)$ vectors ${\bf \cal G}$ and ${\bf g}$

$$\begin{array}{c}
{\bf \cal G}=\left( \begin{array}{c}
                 {\bf \cal G}^0\\
                 {\bf \cal G}^1
                 \end{array}\right),
  {\bf \cal G}^0=\left( \begin{array}{c}
                {\cal G}^{0\nu^\prime}_{0m^\prime}\\
                \vdots\\
                {\cal G}^{0\nu^\prime}_{Nm^\prime}\\
               \end{array}\right),
   {\bf \cal G}^1=\left( \begin{array}{c}
                {\cal G}^{1\nu^\prime}_{0m^\prime}\\
                \vdots\\
                {\cal G}^{1\nu^\prime}_{Nm^\prime}\\
               \end{array}\right),\\
  {\bf g}=\left( \begin{array}{c}
                 {\bf g}^0\\
                 {\bf g}^1
                 \end{array}\right),
  {\bf g}^0=\left( \begin{array}{c}
                {g}^{0\nu^\prime}_{0m^\prime}\\
                \vdots\\
                {g}^{0\nu^\prime}_{Nm^\prime}\\
               \end{array}\right),
   {\bf g}^1=\left( \begin{array}{c}
                {g}^{1\nu^\prime}_{0m^\prime}\\
                \vdots\\
                {g}^{1\nu^\prime}_{Nm^\prime}\\
               \end{array}\right).
\end{array}\eqno{(9)} $$
From Eqs. (6) and (7), we obtain
$${\bf \cal G}=(I-M)^{-1}{\bf g},\eqno{(10)}$$
where $I$ is $2(N+1)\times 2(N+1)$ unit matrix and
$$M=\left[ \begin{array}{cc}
           M^{11}&M^{12}\\
           M^{21}&M^{22}\\
           \end{array}\right]\eqno{(11)}$$
with $M^{11}, M^{12}, M^{21}$ and $M^{22}$ are $(N+1)\times (N+1)$
matrices whose matrix elements are as follows
$$\begin{array}{lll}
M^{11}_{mm^\prime}&=&(a_mf^*_{{\bf k}+m{\bf Q}}+b_mg^*_{{\bf k}+
                 m{\bf Q}})\delta_{m+1 m^\prime}\\
    &    &        +(a_mf_{{\bf k}+(m-1){\bf Q}}+b_mg_{{\bf k}+
                  (m-1){\bf Q}})\delta_{m-1 m^\prime},\\
M^{12}_{mm^\prime}&=&(a_mg^*_{{\bf k}+m{\bf Q}}-b_mf^*_{{\bf k}+
                 m{\bf Q}})\delta_{m+1 m^\prime}\\
    &    &        +(a_mg_{{\bf k}+(m-1){\bf Q}}-b_mf_{{\bf k}+
                  (m-1){\bf Q}})\delta_{m-1 m^\prime},\\
M^{21}_{mm^\prime}&=&(b_mf^*_{{\bf k}+m{\bf Q}}+c_mg^*_{{\bf k}+
                 m{\bf Q}})\delta_{m+1 m^\prime}\\
    &    &        +(b_mf_{{\bf k}+(m-1){\bf Q}}+c_mg_{{\bf k}+
                  (m-1){\bf Q}})\delta_{m-1 m^\prime},\\
M^{22}_{mm^\prime}&=&(b_mg^*_{{\bf k}+m{\bf Q}}-c_mf^*_{{\bf k}+
                 m{\bf Q}})\delta_{m+1 m^\prime}\\
    &    &        +(b_mg_{{\bf k}+(m-1){\bf Q}}-c_mf_{{\bf k}+
                  (m-1){\bf Q}})\delta_{m-1 m^\prime}.\\
\end{array}\eqno{(12)}$$
We note that the charged modulations observed in STM experiments
are weak, i.e. $f_{{\bf k}+m{\bf Q}}$ and $g_{{\bf k}+m{\bf Q}}$
(or $\lambda's$) are small. Expanding Eq. (10), we have ${\bf \cal
G}=(I+M+M^2+\cdots){\bf g}$. From Eq. (7),we finally obtain the
green's functions $$\begin{array}{c} G^{\nu\nu^\prime}_{m
m^\prime}({\bf k}, {\bf
k}^\prime;i\omega_n)=\sum_{\nu^{\prime\prime}}
(-1)^{(1-\nu^{\prime\prime})\nu}\xi_{{\bf k}m\nu+
\nu^{\prime\prime}}\\ ~~~~~~~~\times {\cal
G}^{\nu^{\prime\prime}\nu^\prime}_{mm^\prime} ({\bf k},{\bf
k}^\prime;i\omega_n).\end{array}\eqno{(13)}$$

To compare with the STM experiments, up to the first order in $\lambda'$s,
we derive the local density of states
$$\begin{array}{l}
\rho({\bf r},\omega)=-\frac{1}{\pi}{\rm Im}\sum_\sigma[-{\cal F}
<c_{{\bf r}\sigma}(\tau)c^\dag_{{\bf r}\sigma}(0)>]
|_{i\omega_n\rightarrow\omega+i0^+}\\
~~ =-\frac{2}{{\cal N}\pi}{\rm Im}\sum_{{\bf k}m}
\{a_m({\bf k};i\omega_n)
+(f_{{\bf k}+m{\bf Q}}e^{i{\bf Q} \cdot {\bf r}}+c.c)\\
~~  \times [a_m({\bf k};i\omega_n)a_{m+1}({\bf k};i\omega_n)
-b_m({\bf k};i\omega_n)b_{m+1}({\bf k};i\omega_n)] \\
~~ +(g_{{\bf k}+m{\bf Q}}e^{i{\bf Q} \cdot {\bf r}}+c.c)
 [a_m({\bf k};i\omega_n)b_{m+1}({\bf k};i\omega_n)\\
~~ +b_m({\bf k};i\omega_n)a_{m+1}({\bf k};i\omega_n)]\}|_{
i\omega_n\rightarrow \omega+i0^+},
\end{array}\eqno{(14)}$$
where ${\cal N}$ is the number of sites in the lattice and
$c_{{\bf r}\sigma}={\cal N}^{-1/2}\sum_{{\bf k}m}
c_{{\bf k}+m{\bf Q}\sigma}e^{i({\bf k}+m{\bf Q})\cdot {\bf r}}$.
Obviously, the first term in Eq. (14) is nothing but
LDOS for superconducting state. The other terms are the LDOS
modulations with period $2\pi/{\rm Q}$ due to the hopping
and pairing modulations, respectively. The Fourier component
of LDOS at the ordering wavevector ${\bf Q}$ is
$$\begin{array}{lll}
\rho_{\bf Q}(\omega)&=& \frac{1}{{\cal N}}\sum_{\bf r}
e^{-i{\bf Q}\cdot{\bf r}}\rho({\bf r},\omega)\\
& =&-\frac{2}{{\cal N}\pi}\sum_{{\bf k}m}
\{f_{{\bf k}+m{\bf Q}}{\rm Im} [a_m({\bf k};i\omega_n)a_{m+1}({\bf k};i\omega_n)\\
& &-b_m({\bf k};i\omega_n)b_{m+1}({\bf k};i\omega_n)] \\
& &+g_{{\bf k}+m{\bf Q}} {\rm Im} [a_m({\bf k};i\omega_n)b_{m+1}({\bf k};i\omega_n)\\
& & +b_m({\bf k};i\omega_n)a_{m+1}({\bf k};i\omega_n)]\}|_{
i\omega_n\rightarrow \omega+i0^+}.
\end{array}\eqno{(15)}$$
It is clear that when several of the charge density wave orders
exist simultaneously, the total $\rho_{\bf Q}(\omega)$ is obtained
by a superposition of those of them.

In Fig. 1, we show the real parts of $\rho_{\bf Q}({\omega})$ for
different charge density wave orders in $Bi_2Sr_2CaCu_2O_{8+\delta}$,
whose imaginary parts are zero or are
proportional to its real parts. Obviously, the experimental curves (Fig. 3)
of Ref. [2] cannot be explained by the site- or bond-centered
charge modulation alone (Fig.1a). But our results for
the dimerization hopping (Fig. 1b, c) and transverse pairing modulations
(Fig. 1e) are consistent with the STM experiments. Because the imaginary
part of $\rho_{\bf Q}(\omega)$ observed in the STM experiments is
small, we conclude that the charge stripes are mainly formed by
the transverse dimerization hopping and transverse pairing modulations.
We note that our results are somewhat different from those of Ref. [8],
this may be due to the approximations using in Ref. [8].
We also note that the charge modulations discussed in Ref. [9] are
two-dimensional but ours are one-dimensional.

Fig. 2a-e show Re$\rho_{\bf Q}({\omega})$ for $YBa_2Cu_3O_{6.35}$ with
the same charge density wave orders as in Fig. 1. The curves in Fig. 1
and Fig. 2 are very different due to the different ordering wavevector
and doping in two superconductors. For the dimerization hopping
modulations (Fig. 2 b, c), Re$\rho_{\bf Q}({\omega})$ has single
peak rather than two peaks. For the pairing modulations (Fig. 2 d, e),
there exist a strong peak and a weak peak at symmetric positions.
These results are expected to be verified by the STM experiments.

Finally, we discuss the ARPES experiments on the charge stripes,
which measure the spectral function.
From Eqs. (3) and (13), we obtain the spectral function
up to the second order correction
$$\begin{array}{l}
A_{\bf K}(\omega)\equiv A_m({\bf k},\omega)\\
=-\frac{1}{\pi}{\rm Im}[-{\cal F}
<c_{{\bf k}+m{\bf Q}\uparrow}(\tau)c^\dag_{{\bf k}+m{\bf Q}\uparrow}(0)>]
|_{i\omega_n\rightarrow\omega+i0^+}\\
=-\frac{1}{\pi}{\rm Im}a_m({\bf k};i\omega_n)\{1+\sum_\nu
[a_{m+\nu-1}({\bf k};i\omega_n)f^*_{{\bf k}+(m+\nu-1){\bf Q}}\\
+b_{m+\nu-1}({\bf k};i\omega_n)g^*_{{\bf k}+(m+\nu-1){\bf Q}}]
[a_{m+\nu}({\bf k};i\omega_n)f_{{\bf k}+(m+\nu-1){\bf Q}}\\
+b_{m+\nu}({\bf k};i\omega_n)g_{{\bf k}+(m+\nu-1){\bf Q}}]
+[b_{m-1}({\bf k};i\omega_n)f^*_{{\bf k}+(m-1){\bf Q}}\\
+c_{m-1}({\bf k};i\omega_n)g^*_{{\bf k}+(m-1){\bf Q}}]
[a_{m}({\bf k};i\omega_n)g_{{\bf k}+(m-1){\bf Q}}\\
-b_{m}({\bf k};i\omega_n)f_{{\bf k}+(m-1){\bf Q}}]
+[b_{m+1}({\bf k};i\omega_n)f_{{\bf k}+m{\bf Q}}\\
+c_{m+1}({\bf k};i\omega_n)g_{{\bf k}+m{\bf Q}}]
[a_{m}({\bf k};i\omega_n)g^*_{{\bf k}+m{\bf Q}}\\
-b_{m}({\bf k};i\omega_n)f^*_{{\bf k}+m{\bf Q}}]\}
|_{i\omega_n\rightarrow\omega+i0^+}\\
\equiv A_{0\bf K}(\omega)+\delta A_{\bf K}(\omega)
\end{array}\eqno{(16)}$$
The first term in the above equation is the spectral
function of the superconducting state. The others are
those due to the weak charge modulations, which are the second
order small quantities. Obviously, the spectral functions
$A_{\bf K}(\omega)$ for different hopping (pairing) modulations
have similar energy dependence at the same momentum ${\bf K}$.
So it is difficult to clarify which kinds of hopping (pairing)
modulations  the charged stripes belong by the ARPES experiments.
However, the hopping and pairing modulations have unique
energy dependence for $A_{\bf K}(\omega)$ which could be
distinguished experimentally. Fig. 3 and Fig. 4 show the spectral
functions $A_{\bf K}(\omega)$ at various momenta ${\bf K}$
for $Bi_2Sr_2CaCu_2O_{8+\delta}$
and $YBa_2Cu_3O_{6.35}$ with hopping and pairing modulations, respectively.

In summary, we study the effects of various charge density wave
orders on the d-wave superconductor. We conclude that the dimerization
hopping and transverse pairing modulations and experimental results
are in excellent agreement. The origin of such charge modulations
is due to the hopping and gap disorders in the superconductor, which
will be investigated in detail in another paper [12]. We also discuss
the ARPES experiments on the charged stripes, which can distinguish
the hopping and pairing modulations.

The author wish to thank Prof. C. S. Ting for useful discussions.
This work has been supported by the Texas Center for
Superconductivity at the University of Houston and by the Robert
A. Welch Foundation.

\begin{figure}
\narrowtext \caption {Energy dependence
Re$\rho_{\bf Q}(\omega)$ of the Fourier component of the LDOS at
${\bf Q}=(\pi/2,0)$ for $Bi_2Sr_2CaCu_2O_{8+\delta}$ with
different charge density wave orders. (a) A
site- or bond-centered charge density wave. (b) A longitudinal dimerization
charge density wave. (c)  A transverse dimerization charge density wave.
(d)  A longitudinal pairing modulation. (e) A transverse pairing modulation.
In both (b) and (d), Re$\rho_{\bf Q}(\omega)$ was multiplied by -1 to compare
conveniently with the experiments.  }
\end{figure}

\begin{figure}
\narrowtext \caption {Re$\rho_{\bf Q}(
\omega)$ at ${\bf Q}=(\pi/4,0)$ for $YBa_2Cu_3O_{6.35}$ with the same
charge density wave orders as in Fig. 1.}
\end{figure}

\begin{figure}
\narrowtext \caption {Spectral function
$A_{\bf K}(\omega)$ vs $\omega$ at various ${\bf K}$ for
$Bi_2Sr_2CaCu_2O_{8+\delta}$ with hopping and pairing modulations.}
\end{figure}

\begin{figure}
\narrowtext \caption {Spectral function
$A_{\bf K}(\omega)$ vs $\omega$ at various ${\bf K}$ for
$YBa_2Cu_3O_{6.35}$ with hopping and pairing modulations.}
\end{figure}

\end{multicols}

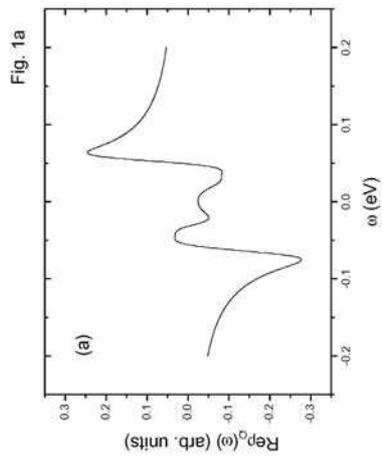
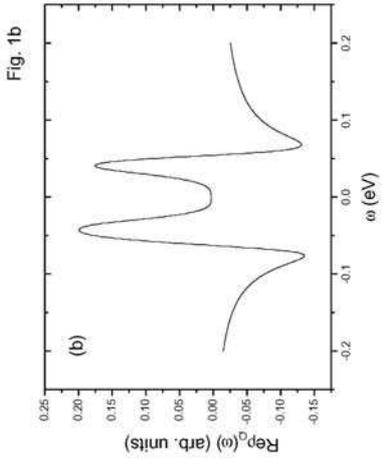
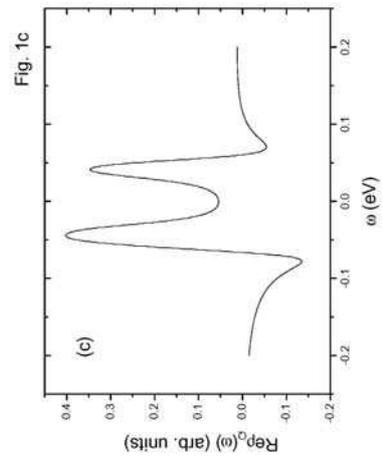
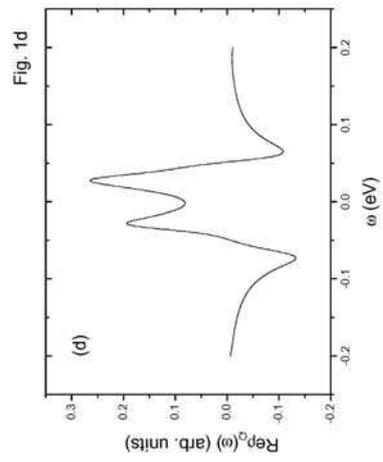
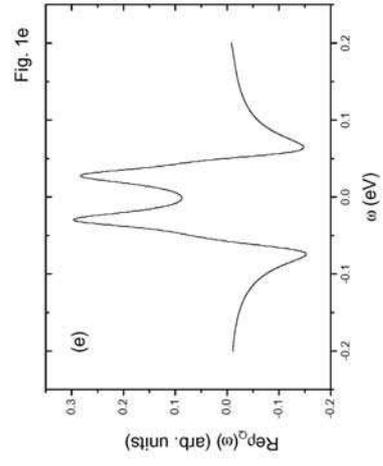
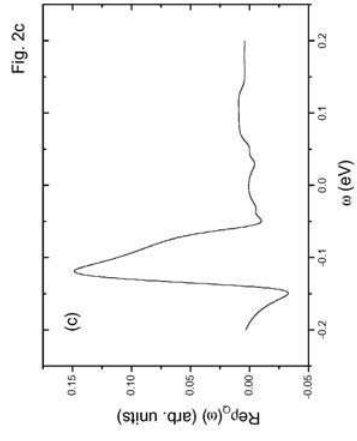
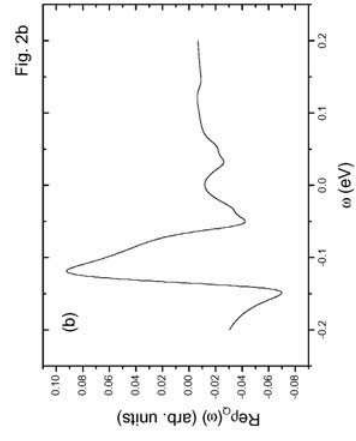
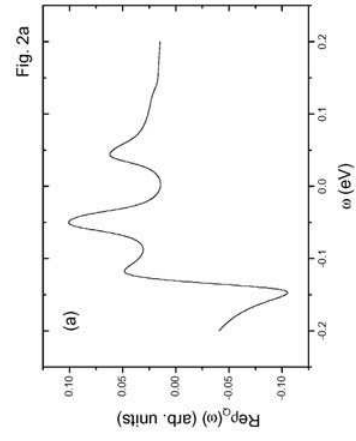
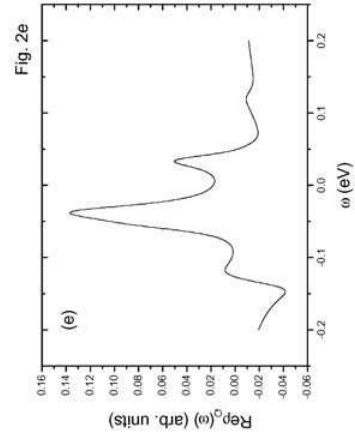
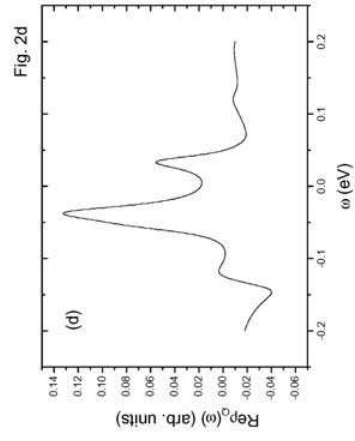
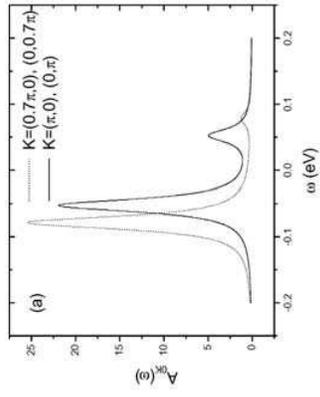
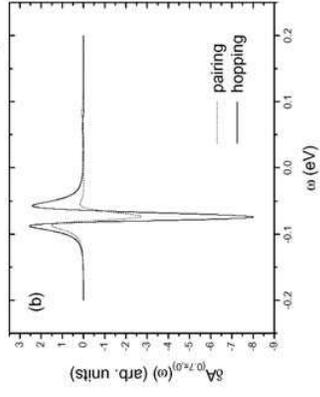
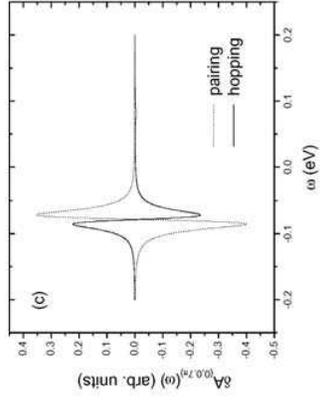
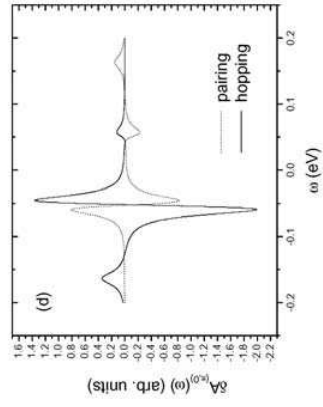
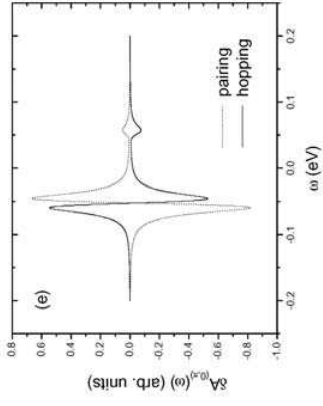
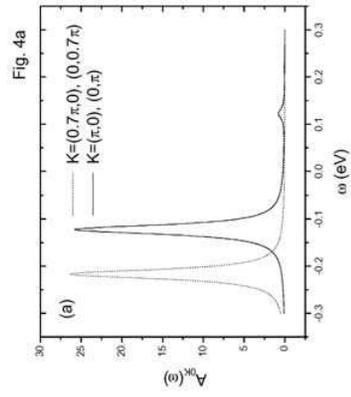
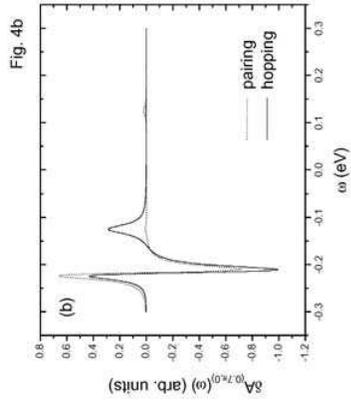
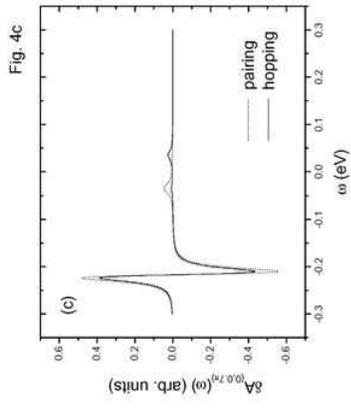
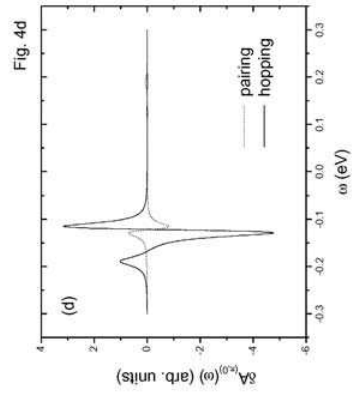
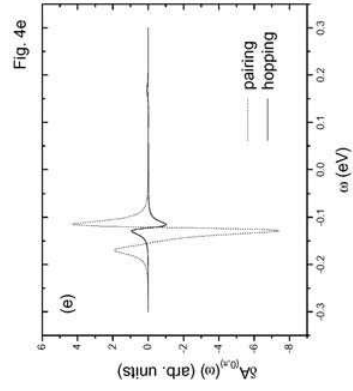
\end{document}